\documentclass[onecolumn,amsmath,amssymb,prd,reprint,longbibliography,floatfix,8pt]{revtex4-1}

\usepackage{cancel}
\usepackage{enumitem}
\usepackage[utf8x]{inputenc}
\usepackage{graphicx}
\usepackage{dcolumn}
\usepackage{braket}
\usepackage{bm}
\usepackage{comment}
\usepackage[switch]{lineno}
\usepackage{float} 
\usepackage{subfigure}
\usepackage{tikz}
\usepackage{multirow}
\usetikzlibrary{arrows.meta}
\usetikzlibrary{arrows,decorations.pathmorphing,backgrounds,positioning,fit,petri}

\begin{document}

\title{Triangulene-based diradicals as a blueprint for molecular quantum platforms \\ with optical addressability and long spin coherence times}

\author{Arup Sarkar}
\author{Cathal Hogan}
\author{Conor Ryan}
\author{Lorenzo A. Mariano}
\author{Alessandro Lunghi}
\email{lunghia@tcd.ie}
\affiliation{School of Physics, CRANN Institute, and RINN Quantum, Trinity College, Dublin 2, Ireland}

\begin{abstract}
{The identification of molecules that combine long spin coherence times and efficient spin-optical interfaces, ideally at room temperature, is pivotal towards the development of molecular quantum technology. By means of advanced first-principles methods, we here unravel the electronic structure for triangulene \textbf{(1)}, its aza-cation derivative \textbf{(2)}, and the crystal of 2,6,10-tri-tert-butyl-4,8,12-trimesityl-triangulene \textbf{(3)}, and show that these organic diradicals possess a triplet ground state well separated from the first singlet excited state approaching 0.5 eV, closely resembling solid-state defects like nitrogen vacancy centers. In addition, we compute spin decoherence times due to the interaction with phonons and surrounding nuclear spins, showing that a deuterated molecule of \textbf{3} in a nuclear spin-free environment would support $T_2 = 0.21$ ms at 10 K. Importantly, we show that the engineering of specific low-energy vibrations could significantly improve $T_2$ toward the limit imposed by the molecular core spin relaxation, here estimated to be as long as $T_1=27$ ms at 300 K for \textbf{2}. Finally, we compute two-phonon contributions to inter-system crossing at 300 K for \textbf{2} as a luminescent prototype, and find that it is highly spin-selective, supporting the possibility to engineer optical read out and spin initialization. These results advance a unified first-principles theoretical foundation of spin decoherence and spin-selective excited-state processes and point to novel chemical design strategies for optically addressable, highly coherent molecular qubits.}
\end{abstract}

\maketitle

\section*{Introduction}

The experimental observation that magnetic molecules can exhibit long spin coherence times\cite{PhysRevLett.98.057201} has sparked a large interest in the transformative role they could play in quantum information science and technology, for instance, offering a practical avenue toward nanometric quantum sensors \cite{Yu2021} or self-error-correcting computing qubits.\cite{Chiesa2024MolecularProcessing} A central challenge in this field is represented by the design of molecular units, or qubits, able to exhibit multiple optimal quantum functionalities at the same time.\cite{Atzori2019TheChemistry, Wasielewski2020ExploitingScience} For instance, many efforts have been devoted to further increase the resilience of spin coherence against both magnetic and thermal noise, and several chemical or spectroscopic strategies have been used to successfully achieve coherence times approaching ms and $\mu$s at cryogenic and room temperature ($T$), respectively.\cite{Zadrozny2015MillisecondQubit, Atzori2016QuantumMoiety, Shiddiq2016EnhancingTransitions} However, performing quantum operations on molecular spin qubits also requires i) initializing their spin in a well-defined, reproducible and controlled pure quantum state, and ii) reading out said state as a function of time.\cite{Atzori2019TheChemistry}  All this needs to be achieved at the single-molecule level, and, depending on the specific application, at high temperature, e.g. ideally room $T$ for biological sensing.\cite{Ishiwata2026MolecularCells} These are, of course, enormous challenges for the conventional magnetic resonance setups generally used to measure molecular spin coherence. The natural energy separation between spin sublevels remains extremely small even in the presence of high magnetic fields, requiring cryogenic temperatures to achieve spin initialization. Similarly, conventional electron paramagnetic resonance is only able to detect $\sim$ 10$^9$ spins over 1 s of acquisition time, making the readout of single molecules an outstanding challenge.\cite{Artzi2015Induction-detectionSpins}

Alternative ways to control spins have also been pursued. For instance, single-molecule electric read-out has been achieved in STM or break junctions.\cite{Vincent2012ElectronicTransistor, Baumann2015ElectronSurface} These experiments have been used to deliver landmark demonstrations of the viability of magnetic molecules for quantum technology, such as the implementation of Grover's search algorithm\cite{PhysRevLett.119.187702} or the sensing of nearby atomic magnetic impurities,\cite{Reale2024ElectricallySurface} but suffer from very restrictive operating conditions, e.g. mK temperatures, and inherent fragility. The use of optical means represents an exciting alternative, albeit not free from challenges. The earliest examples of the use of light to read out and initialize molecular spins in the excited triplet states of arenes date back as far as the 60s,\cite{Kwiram1967OpticalStates} with the observation of single-molecule optically detected magnetic resonance (ODMR) at low temperatures achieved in pentacene in the early 90s.\cite{Kohler1993MagneticSpin} Renewed interest in this molecule has recently been shown, and advanced optical control schemes have been performed in ensembles up to room temperature.\cite{Mann2025ChemicallyResonance} Alternative strategies have also been pursued. For instance, ground-state spin initialization has been achieved in compounds with ultra-narrow absorption linewidths by selectively exciting transitions involving one spin orientation and therefore pumping population into those spin states not affected by absorption.\cite{Bayliss2020OpticallyProcessing}The use of circularly polarized light to initialize spin and Faraday rotation to optically read it out has also been recently proposed.\cite{Sutcliffe2024UltrafastSolution} All of these attempts represent outstanding proof-of-concept that optical spin control can be achieved, but all suffer from their own limitations, and room-temperature, single-molecule, highly coherent spin control still remains to be achieved.\cite{Atzori2025OpticalMolecules}

\begin{figure*}[!ht]
  \centering
  \vspace{5pt}
  \includegraphics[width=1.0\textwidth, trim=0 0 0 0, clip]{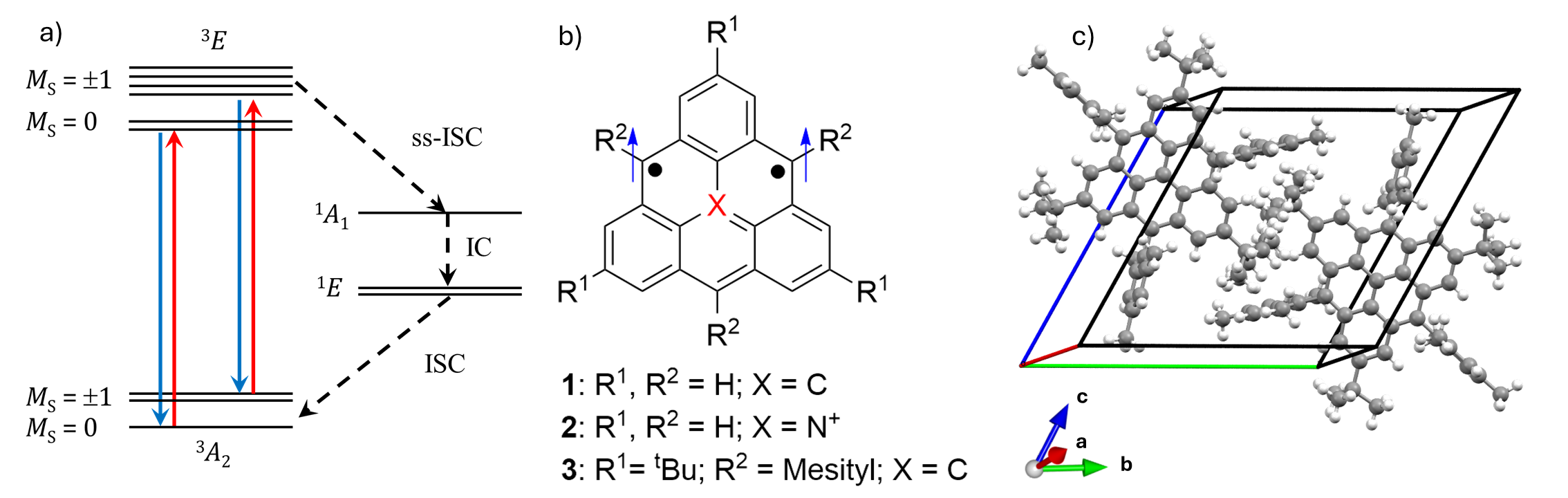}
  \caption{ \textbf{Schematics of electronic processes and triangulene-based diradical structures.} a) Schematic representation of the transitions involved in an NV$^{-}$ center's optical cycle in zero magnetic field. Dashed lines represent ISC and Internal Conversion (IC). Continuous red and blue lines represent spin-preserving photon absorption and emission. States are labelled according to defect's C$_{3v}$ point-group symmetry. b) Molecular representation of Compounds \textbf{1}, \textbf{2} and \textbf{3}. c) The experimental crystal structure of compound \textbf{3}.}
  \vspace{-5pt}
  \label{Fig1}
\end{figure*}
Interestingly, if we move away from the molecular world, there are known examples of solid-state defects that possess efficient room-temperature spin-optical interfaces and long coherence times.\cite{Wolfowicz2021QuantumDefectsb} The prototypical example for this is the spin of negative nitrogen-vacancy defects in diamond (NV$^-$ centers), which can be singularly optically initialized and read-out at ambient temperature. Thanks to their very peculiar electronic structure,\cite{Gali2019AbDiamond} represented schematically in Fig. \ref{Fig1}a, NV$^-$ centers can undergo an inter-system crossing (ISC) event to a singlet state from their optically accessible triplet excited state. Key to the optical properties of NV$^-$ centers, this ISC event is spin selective (ss-ISC) and favors the states $M_s=\pm 1$. Therefore, as the population is promoted to the excited triplet state by laser irradiation, the $M_s=0$ component has a higher probability to fluoresce than the states $M_s=\pm 1$. As a consequence, the luminescence of the NV$^-$ center is proportional to the population with $M_s=0$, which can be used to read out the spin states by ODMR. On the other hand, due to a loss of spin selectivity for the reverse ISC process ($^1E \: \rightarrow \: ^3A$) the states with $M_s=\pm 1$ will have a finite probability to thermalize into any of the three $M_s$ components of the triplet ground state. Continuous laser excitation will thus eventually convert a distribution of states into the $M_s=0$ component, achieving spin initialization. In addition, the fast rate of these electronic processes (ns) over $T_2$ timescales (ms) and the large quantum yield of NV$^{-}$($\sim 30\%$) make the optical cycle very efficient.\cite{Wolfowicz2021QuantumDefectsb}

A theoretical proposal to reproduce NV centers' electronic structure at the molecular level has been presented,\cite{Poh2024AlternantQubits} and a few experimental attempts have shown high promise. For instance, a luminescent triplet diradical based on the latter proposal has been shown to exhibit an ODMR contrast of 1\% at 5 K, and quantum yields of $\sim$1\%.\cite{Kopp2024LuminescentQubits} Shortly after, low-$T$ ODMR has also been achieved in highly luminescent diradicals, but at the cost of losing the triplet ground state due to antiferromagnetic diradical interactions.\cite{Chowdhury2025BrightStates} Further molecular design appears to be capable of tackling some of these limitations, with the engineering of excited states\cite{Poh2025EnhancingQubits}, diradical bridges\cite{Kopp2025OpticallyQubits,Chowdhury2025RoomDiradicals}, and host matrices\cite{Kopp2026MolecularSensing} appearing as promising ways forward. In addition, the use of carbene\cite{Poh2025ElectronCarbenes} and nitrene\cite{Ricci2025OpticalTriplet} diradicals has been proposed to overcome these initial limitations in stabilizing a high-spin ground state, and early experimental support at cryogenic temperatures is already present in the literature.\cite{Roggors2025OpticallyQubits, Roggors2026AInterface} Overall, this early literature suggests that replicating NV centers' electronic structure and photophysics might well be possible. However, while the gap is rapidly shrinking, none of the current proposals appears to simultaneously embed all the elements that make NV$^{-}$ centers so unique. These include a well-isolated, highly coherent triplet ground state, fast and high luminescence, high chemical stability, and highly ss-ISC for ODMR contrast and full spin initialization. We argue that achieving a full matching of defects' spin-optical interfaces efficiency is likely to take exploring innovative chemical solutions, as well as deepening our understanding of spin-selective electronic dynamics to guide these efforts. Indeed, while ab initio simulations are already playing a role in this space,\cite{Poh2024AlternantQubits, Poh2025ElectronCarbenes, Ricci2025OpticalTriplet, Poh2025EnhancingQubits} they have not yet tackled the description of either ground- or excited-states spin-resolved spin dynamics, leaving the exact mechanism of ss-ISC and decoherence unknown.

In this work, we address both of these challenges by advancing the unification of first-principles theories of spin decoherence and excited states relaxation, and exploring the possibility of using triangulene (also known as Clar's hydrocarbon) as a template to construct molecules with efficient spin-optical interfaces and long spin coherence times. Triangulene had been theorized to have a non-Kekule structure and a triplet ground state\cite{Inoue2001TheHydrocarbon}, but its high instability had only allowed a partial characterization at low temperature for a long time. A breakthrough in this area was achieved by demonstrating its high stability through on-surface characterization, an observation closely followed by the development of several families of open-shell graphenoides with different topological and exchange structures.\cite{Pavlicek2017SynthesisTriangulene} Through a proper chemical functionalization of its scaffold, triangulene has also been stabilized in solid-state and solution, and several derivatives have been synthesized and fully characterized in recent years.\cite{Inoue2001TheHydrocarbon, Morita2011SyntheticFragments, Arikawa2021SynthesisTriangulene, Arikawa2023AState} Magnetic measurements confirmed its triplet ground state and optical spectroscopy measurements confirm the presence of well-separated singlet states,\cite{Pavlicek2017SynthesisTriangulene} similar to NV centers, but a full understanding of its electronic structure and dynamical properties is currently lacking. Here we employ multireference quantum chemistry simulations to determine the electronic properties of triangulene (\textbf{1}) and its derivative aza-triangulene cation (\textbf{2}), depicted in Fig. \ref{Fig1}b, and show that these molecules exhibit an electronic structure strikingly similar to that of an NV center. In addition to this, we expand the scope of multireference relativistic spin-vibronic coupling and spin relaxation simulations\cite{Mariano2024Spin-VibronicFormalism, Mariano2025TheMolecules} to organic open-shell molecules and predict that \textbf{1} and \textbf{2} should both exhibit long spin-phonon relaxation times at room temperature. The extension of our study to a known crystal of a kinetically stabilized triangulene derivative, 2,6,10-tri-tert-butyl-4,8,12-trimesityl-triangulene (\textbf{3}), \cite{Arikawa2021SynthesisTriangulene} shows that the chemical functionalization and lattice environment of the diradical scaffold play a key role in modulating triangulene's spin relaxation and dephasing rates,\cite{Ryan2025SpinBaths, Lunghi2026AMolecules} suggesting possible avenues to preserve long coherence times at room temperature of \textbf{1}-\textbf{2} in realistic solid-state matrices. Finally, we use the same formalism to compute two-phonon vibronic transitions\cite{Mariano2025TheMolecules} to excited states relaxation, predicting that highly ss-ISC processes are operative at room temperature, possibly supporting ground-state optical read out and spin initialization. On the one hand, these results provide strong theoretical evidence that triangulene-based diradicals offer a chemically tunable and flexible architecture for the development of molecular quantum systems with optical addressability and long coherence times. On the other hand, the full decomposition of these results into first principles and the steps taken to establish a unified ab initio theory of spin and excited states dynamics in open shell systems provide a general roadmap for the design and optimization of novel molecules with advanced quantum properties.

\section*{Theoretical methods}

\textbf{Relativistic vibronic coupling.} The complete Hamiltonian describing the open quantum system of electrons interacting with nuclei can be expressed as\cite{Mariano2025TheMolecules}
\begin{equation}
\hat{H} = \hat{H}_{\mathrm{e}} + \hat{H}_{\mathrm{ph}} + \hat{H}_{\mathrm{e-ph}} \:.
\label{eq:1}
\end{equation}

The Born-Oppenheimer electronic Hamiltonian, 
\begin{equation}
   \hat{H}_{\rm e} = \hat{H}_{0} + \hat{H}_{\rm soc} + \hat{H}_{\rm ssc} + \hat{H}_{\rm z} \:,
\end{equation}
includes both the non-relativistic spin-free kinetic and coulomb term, $\hat{H}_{0}$, and the Hamiltonians describing spin-orbit, $\hat{H}_{\rm soc}$, spin-spin, $\hat{H}_{\rm ssc}$, and Zeeman, $\hat{H}_{\rm z}$, relativistic interactions. The nuclear dynamics for each individual potential energy surface is described at the harmonic level through the Hamiltonian
\begin{equation}
\hat{H}_{\mathrm{ph}} =
\sum_{\alpha} \hbar \omega_{\alpha} \left( \hat{n}_{\alpha} + \frac{1}{2} \right) \:,
\label{eq:2}
\end{equation}
where $\hbar \omega_{\alpha}$ is the energy quanta of the phonon $Q_{\alpha}$ and $\hat{n}_{\alpha}$ is the operator describing the number of such phonons. Here, we assume that every electronic state shares the same set of phonons, usually of the ground state. Finally, the last term in Eq. \ref{eq:1} is the coupling between electrons and phonons, namely the vibronic coupling. The latter is defined as a Taylor expansion of $\hat{H}_{e}$ with respect to the atomic displacements associated with the phonons $Q_{\alpha}$. Considering only up to the quadratic term, this reads
\begin{equation}
\hat{H}_{\mathrm{e-ph}} = \sum_{\alpha} \hat{V}^\alpha \: Q_{\alpha} + \frac{1}{2}\sum_{\alpha\beta} \hat{V}^{\alpha\beta} \: Q_{\alpha} Q_{\beta} \:,
\label{eq:3}
\end{equation}
where
\begin{equation}
\hat{V}^\alpha = \left( \frac{\partial \hat{H}_{\mathrm{e}}}{\partial Q_{\alpha}} \right)  \:, \quad \text{and} \quad \hat{V}^{\alpha\beta} = \left( \frac{\partial^2 \hat{H}_{\mathrm{e}}}{\partial Q_{\alpha}\partial Q_{\beta}} \right) \:. 
\label{eq:3}
\end{equation}
Focusing on the linear terms, $\hat{V}_\alpha$, these are computed starting from the corresponding Cartesian derivatives as
\begin{equation}
    \left( \frac{\partial \hat{H}_{\mathrm{e}}}{\partial Q_{\alpha}} \right) = \sum_{st}^{3N_{at}}\sqrt{\frac{\hbar}{2m_s\omega_\alpha}} L_{st,\alpha} \left( \frac{\partial \hat{H}_{\mathrm{e}}}{\partial X_{st}} \right) \:,
\end{equation}
where $L_{st,\alpha}$ are elements of the eigenvectors of the Hessian matrix describing how much the Cartesian degree of freedom $t$ of atom $s$, $X_{st}$, with mass $m_s$, contributes to the $\alpha$-th normal mode. A similar corresponding expression applies to quadratic coupling.\cite{Lunghi2022TowardTheoryc}

\textbf{Ab initio simulation of vibronic coupling.} Let us now unpack the microscopic origin of vibronic coupling in terms of the contributions to $\hat{H}_{\rm e}$. In this contribution, we employ a multireference plus quasi-degenerate perturbation theory approach to solving the Schr\"{o}dinger equation for the full $\hat{H}_{\rm e}$. In this framework, we first determine the solution to the spin-free problem,
\begin{equation}
    \hat{H}_0 | \phi _l\rangle = \mathcal{E}_l | \phi _l\rangle \:,
\end{equation}
through methods based on the Complete Active Space Self Consistent Field (CASSCF) (see ESI for details), and then use this set of states $|\phi_l \rangle$ to build a representation for the relativistic wavefunction, $| \psi_a\rangle = \sum_l U_{al} | \phi _l\rangle $. The coefficients $U_{al}$ are determined according to near-degenerate perturbation theory by diagonalizing the relativistic part of the Hamiltonian,
\begin{equation}
    (\hat{H}_{\rm soc} + \hat{H}_{\rm ssc} + \hat{H}_{\rm z}) | \psi _a\rangle = E_a | \psi _a\rangle \:.
    \label{NDPT}
\end{equation}
In this wavefunction representation, the matrix elements of the vibronic coupling operator can be written as\cite{Mariano2024Spin-VibronicFormalism}
\begin{equation}
 \langle \psi _a | \left ( \frac{\partial \hat{H}_{\mathrm{e}}}{\partial X_{st}} \right ) | \psi_b \rangle = J^{st}_{ab} + K^{st}_{ab} \:,
 \label{JK}
\end{equation}
where, according to the generalized Hellman-Feynman theorem, 
\begin{equation}
 J^{st}_{ab} = (E_a - E_b ) \sum_{lm} \: U_{la}^* U_{bm} \: \langle \phi_l | \frac{\partial}{\partial X_{st}}  | \phi_m \rangle \:, 
 \label{Jmat}
\end{equation}
and
\begin{equation}
    K^{st}_{ab} = (E_a - E_b ) \sum_l U^*_{la}\nabla_{st}U_{lb} \:.
    \label{Umat}
\end{equation}
Both the derivatives of the matrix elements $U$ and the spin-free wavefunction $| \phi \rangle$ are determined through numerical differentiation. For a two-step derivative of the spin-free wavefunction, this reads
\begin{equation}
\langle \phi_l | \frac{\partial}{\partial X_{st}} | \phi_m \rangle =  \frac{\langle  \phi_l | \phi_m^{+\delta} \rangle - \langle  \phi_l | \phi_m^{-\delta}\rangle }{2\delta}\:.
\label{HFT}
\end{equation}
In Eq. \ref{HFT}, $| \phi_l^{\pm\delta}\rangle$ represents a spin-free wavefunction computed at a geometry with the degree of freedom $st$ displaced by $\pm\delta$. The wavefunctions overlap $\langle \phi_l | \phi_m^{\pm\delta}\rangle $ are computed through existing numerical recipes.\cite{Plasser2016EfficientOverlaps} Here we also test a multi-step differentiation strategy where $| \phi_l^{\delta}\rangle$ is fitted with a second- or third-order polynomial function of $\delta$ and the linear coefficient is retained. For consistency with the assumption of each electronic state sharing the exact same potential energy surface, we set to zero the diagonal elements of vibronic coupling in the final representation of Eq. \ref{NDPT}, i.e. $V^{\alpha}_{aa}=0$.

In principle, we could use a similar strategy to compute quadratic coupling, but, to the best of our knowledge, this has not yet been implemented and benchmarked. Here, we limit ourselves to computing quadratic coupling for the ground-state multiplet of molecules \textbf{1}-\textbf{3}. In this case, we first map the electronic Hamiltonian $\hat{H}_e$ to a spin Hamiltonian $\hat{H}_s=\vec{\mathbf{S}}\cdot \mathbf{D}\cdot\vec{\mathbf{S}}$,\cite{Mariano2024Spin-VibronicFormalism} where $S=1$ is the effective spin of the ground-state triplet, and then perform a four-step second-order numerical 
differentiation of the zero-field splitting tensor namely $(\partial^2 \mathbf{D}/ \partial X_{st} \partial X_{rv})$.\cite{Lunghi2022TowardTheoryc}

\textbf{Vibronic dynamics.} From the knowledge of the Hamiltonian in Eq. \ref{eq:1} and its matrix elements in the final representation of Eq. \ref{NDPT}, it is possible to determine the dynamics of the electronic system under the influence of the phonons. Building on previous literature,\cite{Lunghi2022TowardTheoryc} we treat the phonons as an equilibrium thermal Markovian bath and use quantum master equations to describe vibronic dynamics. We model one-phonon transitions among electronic states, $a \rightarrow b$, as
\begin{equation}
W^{1\text{-ph}}_{ba} =
\frac{2\pi}{\hbar^{2}}
\sum_{\alpha}
\left| V^{\alpha}_{ba} \right|^{2}
G^{\rm 1-ph}(\omega_{ba}, \omega_{\alpha}, \omega_{\beta}) \:,
\label{W1ph}
\end{equation}
where $V^{\alpha}_{ba}$ are the matrix elements of the linear vibronic coupling coefficients in Eq. \ref{eq:3}, while the function
\begin{equation}
 G^{\rm 1-ph}(\omega_{ba}, \omega_{\alpha})=\bar{n}_\alpha \delta(\omega_{ba}-\omega_\alpha) + (\bar{n}_\alpha+1) \delta(\omega_{ba}+\omega_\alpha)\:,
 \label{Gfunction1}
\end{equation}
introduces energy conservation and phonons' thermal population contributions to phonon absorption and emission, respectively. 
In addition to this, two-phonon transitions account for three possible types of processes: i) the absorption of two phonons, ii) the emission of two phonons, and iii) the absorption of one phonon and the emission of a second one. 
Within the linear vibronic coupling approximation, the transition rates for the latter process emerge at the fourth order of perturbation theory\cite{Mariano2025TheMolecules} and read 
\begin{equation}
W^{2\text{-ph}}_{ba} =
\frac{2\pi}{\hbar^{2}}
\sum_{\alpha > \beta}
\left|
T^{\alpha\beta, +}_{ba}
+
T^{\beta\alpha,-}_{ba}
\right|^{2}
G^{\rm 2-ph}(\omega_{ba}, \omega_{\alpha}, \omega_{\beta}) \:,
\label{eq:5}
\end{equation}
where
\begin{equation}
T^{\alpha\beta,\pm}_{ba} = \sum_{c} \frac{ V^{\alpha}_{bc} V^{\beta}_{ca} }{ E_{c} - E_{a} \pm \hbar \omega_{\beta} }\:.
\label{eq:6}
\end{equation}
Here, it is important to note that all electronic excited states contribute to the computed rate for a specific transition by supporting virtual processes mediated by vibronic coupling, as mathematically described by the sum over all electronic states $c$ appearing in Eq. \ref{eq:6}. The function $G^{\rm 2-ph}$ accounts for the phonons' thermal population and imposes energy conservation, and reads
\begin{equation}
 G^{\rm 2-ph}(\omega_{ba}, \omega_{\alpha}, \omega_{\beta})=\bar{n}_\alpha(\bar{n}_\beta+1) \delta(\omega_{ba}-\omega_\alpha + \omega_\beta)\:.
 \label{Gfunction}
\end{equation}
In addition, quadratic coupling introduces an additional source of two-phonon transitions at the second order of perturbation theory,\cite{Lunghi2022TowardTheoryc} described by
\begin{equation}
W^{2\text{-ph}}_{ba} =
\frac{2\pi}{\hbar^{2}}
\sum_{\alpha > \beta}
\left| V^{\alpha\beta}_{ba} \right|^{2}
G^{\rm 2-ph}(\omega_{ba}, \omega_{\alpha}, \omega_{\beta}) \:,
\label{W2phq}
\end{equation}
where $V^{\alpha\beta}_{ba}$ are the matrix elements of the second-order vibronic coupling coefficients.

Eqs. \ref{eq:5} and \ref{eq:6} can be generalized to describe the evolution of the full density matrix and describe vibrational decoherence,\cite{Lunghi2026AMolecules} inclusive of pure-dephasing contributions. The full expressions for the transition rates for all types of two-phonon processes can be found in the literature.\cite{Lunghi2022TowardTheoryc, Mariano2025TheMolecules} 

\textbf{Spin-spin dephasing.} At low temperatures, vibrational processes become inefficient due to the depletion of phonons' thermal population. In this regime, the dynamics of the nuclear spins surrounding the electron's spin becomes the ultimate limit to decoherence,\cite{PhysRevB.85.115303} assuming extreme magnetic dilution can be achieved.\cite{Ryan2025SpinBaths} Similar to the case of spin-phonon dynamics, this process represents an open quantum system problem and the corresponding Hamiltonian is given by
\begin{equation}
\hat{H} = \hat{H}_{\mathrm{s}} + \hat{H}_{\mathrm{bath}} + \hat{H}_{\mathrm{s-bath}} \:,
\label{Hspinspin}
\end{equation}
where
\begin{align}
    & \hat{H}_s = \vec{\textbf{S}}\cdot\textbf{D}\cdot\vec{\textbf{S}}-\vec{\textbf{B}}\cdot\gamma_S\cdot\vec{\textbf{S}} \:, \label{Hspin} \\
    & \hat{H}_{bath} = -\sum_i \vec{\textbf{B}}\cdot\gamma_i\cdot\vec{\textbf{I}}_i+\sum_{i<j}\vec{\textbf{I}}_i\cdot\textbf{J}_{ij}\cdot\vec{\textbf{I}}_j \:, \label{Hbath} \\
    & \hat{H}_{s-bath} = \sum_i\vec{\textbf{S}}\cdot\textbf{A}_i\cdot\vec{\textbf{I}}_i \:. \label{sbath}
\end{align}
In Eqs. \ref{Hspin}-\ref{sbath}, $\vec{\textbf{S}}$ represents the electron spin of the central molecule, $\{\vec{\textbf{I}}\}$ the spins of the bath, $\textbf{D}$ the zero-field splitting tensor of the central spin, $\textbf{A}$ the hyperfine interaction tensor between the central and bath spins, $\textbf{J}$ the interaction tensor between bath spins, and $\gamma$ the interaction tensor between the spin and the magnetic field (with the convention that $\gamma<0$ for the electron). A quadrupole term $\sum_i\vec{\textbf{I}}_i\cdot\vec{\textbf{P}}_i\cdot\vec{\textbf{I}}_i $ can be included in Eq. \ref{Hbath} in the case the of $I>1/2$ spins.

Due to the mismatch between the gyromagnetic factors of nuclear and electron spins, the Hamiltonian of Eq. \ref{Hspinspin} does not lead to a transfer of energy between the central spin and the spin bath, as is the case for the spin-phonon interactions, but rather contributes to a process called dephasing. In a nutshell, pairs of iso-nuclear spins change their reciprocal orientation with respect to the field (flip-flop), making the effective magnetic field at the electron spin fluctuate, leading to a loss of phase coherence across different spins or quantum measurements and thus to decoherence.

Here, we employ the Cluster Correlation Expansion (CCE) method\cite{Yang2008QuantumBath} as implemented in the software pyCCE\cite{Onizhuk2021PyCCE:Dynamicsc} to simulate these processes. In this approach, the time evolution of the off-diagonal elements of the central spin density matrix describing the spin coherence, $\rho_{ij}$, are computed by studying the dynamics of clusters of spins up to size $M$ by means of the expressions
\begin{equation}\label{eq:rho_M}
    \rho_{ij}^{(M)}(t) = \prod_{|\mathcal{C}|\leq M}\tilde{\rho}_{ij,\mathcal{C}}(t) \:,
\end{equation}
and
\begin{equation}\label{eq:rho_tilde}
    \tilde{\rho}_{ij,\mathcal{C}}(t) = \frac{\rho_{ij,\mathcal{C}}(t)}{\prod_{\mathcal{C'}\subset\mathcal{C}}\tilde{\rho}_{ij,\mathcal{C'}}(t)} \:, 
\end{equation}
where $\rho_{ij,\mathcal{C}}$ is the time evolution of a cluster of spins, $\mathcal{C}$. The latter is computed exactly from unitary evolution as 
\begin{equation}
    \rho_{ij,\mathcal{C}}(t) = \bra{i}e^{-i\hat{H}_\mathcal{C}t/\hbar}(\rho_S(0)\otimes\rho_\mathcal{C}(0))e^{i\hat{H}_\mathcal{C}t/\hbar}\ket{j}\:,
\end{equation}
where the Hamiltonian $\hat{H}_\mathcal{C}$ is obtained by retaining only the terms of  Eq. \ref{Hspinspin} that depend on the spins considered in each specific cluster $\mathcal{C}$. The exact decoherence dynamics is then estimated by showing convergence of $\rho_{ij}^{(M)}(t)$ as a function of $M$.

\section*{Results}

\textbf{Electronic structure.} Triangulene, \textbf{1}, consists of six benzene rings fused together to form a planar triangular aromatic ring with D$_{3h}$ symmetry. The aza-triangulene cation, \textbf{2}, is obtained by substituting the central carbon atom of \textbf{1} with an N$^+$ group, and presents the same symmetry properties. The molecular structure of both molecules is optimized in the gas phase with Density Functional Theory (DFT) assuming a triplet spin multiplicity ground state, as supported by experimental evidence.\cite{Pavlicek2017SynthesisTriangulene} Full details of these simulations are provided as ESI. As is common for conjugated aromatic hydrocarbons, the frontier electronic states can be expected to be composed of $\pi$ and $\pi^*$ excitations arising from the combination of the 22 p$_z$ atomic orbitals of carbon and nitrogen (see Figure S1 in ESI). We investigate the details of such electronic states through multireference simulations based on the CASSCF method and NEVPT2 dynamical correlation corrections.\cite{Angeli2001IntroductionTheory} Full details are reported as ESI. In order to go beyond the computational limits of standard CASSCF implementations, we employed the ICE configuration interaction (ICE-CI) solver\cite{Lang2025TreatingCI} to correlate up to 22 electrons in 19 orbitals. The conclusions of this systematic study, reported in tables S1 and S2 in ESI, highlight two important facts: i) the electronic structure of both \textbf{1} and \textbf{2} is reasonably well described by considering a minimal active space of 10 electrons in 10 orbitals, and ii) dynamical correlation corrections are essential. The remaining discussion is therefore limited to NEVPT2(10,10) results. 

\begin{figure*}[!ht]
    \centering
    \includegraphics[scale=0.5,trim=0cm 14cm 0cm 0cm,clip]{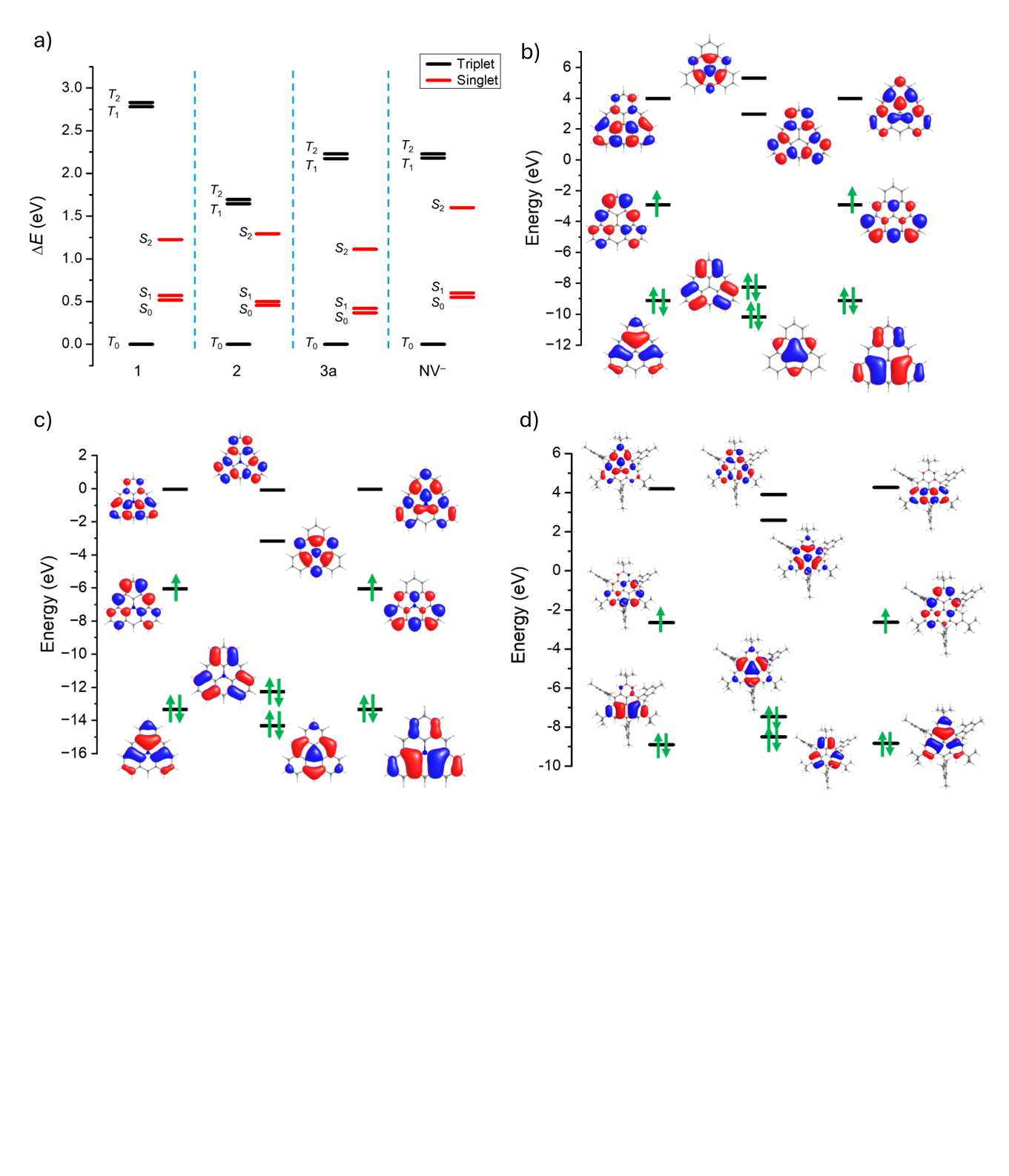}
    \caption{\textbf{Electronic structure of 1-3.} a) Vertical Excitation Energies (VEE) of compounds \textbf{1-3a} and comparison with the experimental VEE of NV$^-$ center in diamond\cite{Haldar2023LocalTheory, Benedek2025AccurateTheory}. The degeneracies of the respective triplets/singlets are artificially lifted to aid visualization. b-d) State average canonical orbitals from CASSCF(10,10) calculations for \textbf{1-3a}. The electronic occupation represents the dominant contribution to the ground-state triplet wavefunction.}
    \label{Figure2}
\end{figure*}

Fig. \ref{Figure2} reports the energy ladder of the first six spin-free electronic states. For both \textbf{1} and \textbf{2}, simulations confirm a triplet ground state ($T_{0,M_S}$) highly stabilized with respect to the first singlet state ($S_{0}$), and the presence of two quasi-degenerate excited triplet states ($T_{1,M_S}$ and $T_{2,M_S}$) above the three singlets (see Fig. \ref{Figure2}a). The most significant difference between \textbf{1} and \textbf{2} lies in the energy of the excited triplets $T_{1,M_S}$ and $T_{2,M_S}$, which are approximately 22,400 cm$^{-1}$ (2.78 eV) and 13,200 cm$^{-1}$ (1.65 eV) above the $T_{0,M_S}$ ground state in \textbf{1} and \textbf{2}, respectively (see Tables S3 and S4). The first two singlets ($S_{0}$ and $S_{1}$) are nearly degenerate and lie around 4,160 cm$^{-1}$ (0.52 eV) and 3,670 cm$^{-1}$ (0.46 eV) above the ground state for \textbf{1} and \textbf{2}, respectively. The third singlet, $S_{2}$, appears at nearly 9,900 cm$^{-1}$ (1.23 eV) and 10,400 cm$^{-1}$ (1.29 eV) above the ground state for \textbf{1} and \textbf{2}, respectively.

To interpret the nature of these many-body electronic states, we report the energy and shape of the underlying canonical Molecular Orbitals (MOs) in Figs. \ref{Figure2}b-d, while their contribution to the total wavefunction is reported as ESI. In both cases, the ground-state MOs occupancy clearly reflects the diradical character. Turning to excited states, key differences between the two molecules arise. For \textbf{1}, the energy separation between the singly occupied MO (SOMO) and doubly occupied MO (DOMO) is comparable to the gap between SOMO and lowest unoccupied MO or LUMO, resulting in a triplet excited state wavefunction with high multireference character and including contributions from both single-electron excitations (see Table S3). In \textbf{2}, however, the electronegativity of N$^+$ has two effects: i) triangulene's LUMO+3 is stabilized greatly due to its higher density on the central atom and becomes the LUMO, ii) the SOMO-LUMO gap shrinks (see Fig. \ref{Figure2}b-c). These facts reflect on the excited triplet wavefunction in \textbf{2} being largely dominated by the single-electron SOMO-LUMO excitation, and explain the much reduced triplet-triplet energy gap in \textbf{2} vs \textbf{1} (see Table S4). The wavefunction analysis for the first two singlet states reveals that one of the near-degenerate singlets is characterized by an open-shell singlet configuration, while the other one is a linear combination of the charge-transfer (spin-paired) closed-shell configurations, involving the SOMOs (see Tables S3 and S4). The third singlet possesses the alternate linear combination of spin-paired configurations.

\begin{figure*}[t]
    \centering
    \includegraphics[scale=1]{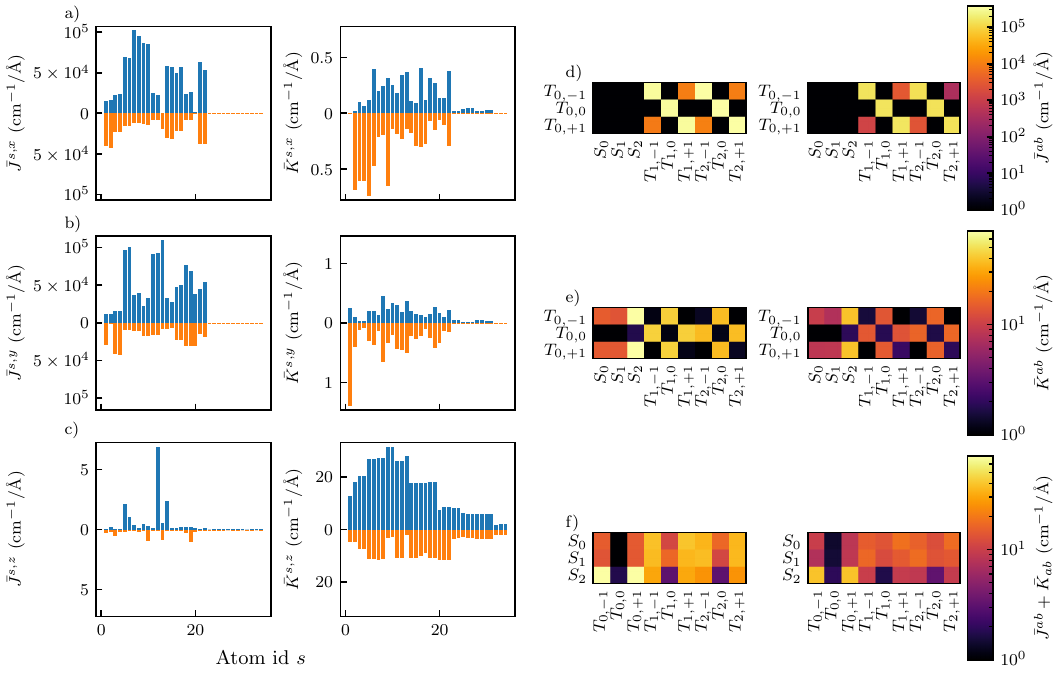}
    \caption{\textbf{Vibronic coupling for 1-2.} Panels a-c) report the decomposition of $J$ and $K$ in atomic and Cartesian contributions. Atom IDs 1-22 correspond to C atoms and 23-34 to H atoms. Values are summed over all possible transitions between any of the three $T_{0,M_S}$ states and all other excited states. Blue (orange) histograms report results for \textbf{1} (\textbf{2}). Panels d-f) report the value of $J$ and $K$ for different transitions after summing over atomic contributions. Left (right) panels report results for \textbf{1} (\textbf{2}).}
    \label{fig3}
\end{figure*}

Let us now turn to the effect of introducing relativistic interactions. In both \textbf{1} and \textbf{2} spin-orbit coupling is observe to preferentially couple only the $M_S=0$ components of the excited triplets, $T_{1,0}$ and $T_{2,0}$, with the $S_{0}$ and $S_{1}$ states, and the ground-state triplet $T_{0,0}$ with the third excited singlet $S_{2}$. Spin-spin interactions instead couple the ground and excited triplet states with $\Delta M_S=\pm 2$. A distinctive feature of this interaction is the loss of degeneracy among different $M_S$ components of the triplet ground state, $T_{0,M_S}$. For the latter, being non-degenerate at the electronic level, we can quantify its zero-field splitting with the spin Hamiltonian
\begin{equation}
    \hat{H}_S = D\hat{S}_z^2 + E (\hat{S}_x^2 - \hat{S}_y^2) \:,
\end{equation}
where $D=0.008$ cm$^{-1}$ and $E/D = 0.0$ for \textbf{1} and $D=0.0094$ cm$^{-1}$ and $E/D = 0.043$ for \textbf{2}. These values are in close agreement with the experimental \textit{D} values reported for triangulene and other tri-substituted derivatives.\cite{Inoue2001TheHydrocarbon, Allinson1993ESRTrioxytriangulene} We note that these values of zero-field splitting are particularly small even for organic molecules, and we attribute it to the quenching of spin-spin interactions due to the large delocalization of the diradical ground-state wavefunction. In addition, spin-spin interactions couple all of the $M_S=\pm1$ excited triplets among themselves. Despite the general weak nature of spin-orbit and spin-spin coupling in organic molecules and the wide energy gaps between the spin-free states of \textbf{1} and \textbf{2}, these interactions play a pivotal role in shaping vibronic coupling and electronic dynamics (\textit{vide infra}).

Compound \textbf{3}, expectedly, has an electronic structure similar to \textbf{1} and \textbf{2} (see Figures S2 and S3). Due to the crystal packing and bulky substituent groups, the overall symmetry of the triangulene ring is diminished in the solid-phase optimized geometry \textbf{3a}. This is also reflected in $S_{0-1}$ and $T_{1-2,M_S}$ now being separated by nearly 200 cm$^{-1}$ (see Table S5). The near degeneracies are somewhat restored for the models \textbf{3b} and \textbf{3c} where the compound \textbf{3} is optimized in the gas phase with CP2K and ORCA packages, respectively (see Tables S6 and S7). The energy of the first two singlets appears at around 3000 cm$^{-1}$ (0.38 eV) in \textbf{3a-c}, comparatively lower than compounds \textbf{1} and \textbf{2} (see Tables S5-S7). Noticeably, the two excited triplet states in compounds \textbf{3a-c} lie at around 17500 cm$^{-1}$ (2.16 eV), midway between the triplets of compounds \textbf{1} and \textbf{2}, and the energies of these two triplets match very closely to the NV$^{-}$ center in diamond (see Fig.~\ref{Figure2}a). Furthermore, the experimental absorption bands\cite{Arikawa2021SynthesisTriangulene} at 533 and 580 nm (17241 and 18761 cm$^{-1}$) match quite well with our NEVPT2(10,10) excitation energies (see Tables S5-S7).

\textbf{Vibronic coupling.} Now that we have presented the electronic structure at the equilibrium geometry for \textbf{1}-\textbf{3}, we are ready to tackle the question of how nuclear motion couples these states through vibronic coupling. All subsequent simulations are performed at a field of 0.33 T, which orders in energy the $M_S$ components of each triplet as -1, 0 and 1. Figs. \ref{fig3}a-c reports the averaged vibronic coupling between the ground state triplet and the excited states for \textbf{1} and \textbf{2} split into its contributions $\bar{J}^{st}$ and $\bar{K}^{st}$ (see Eq. \ref{JK}). The latter are resolved by atom index $s$ and Cartesian degree of freedom $t$. Regarding the physical nature of these terms, $J$ corresponds to the derivatives of the spin-free Hamiltonian rotated in the final wave function representation, therefore representing the contribution of how the Coulomb interactions reshape the wavefunction upon nuclear displacements plus the effect of spin-mixing of states introduced by equilibrium relativistic interactions. $K$ instead represents the direct modulation of relativistic interactions by nuclear displacements. Consistent with this, $\bar{J}^{st}$ is found to be orders of magnitude larger than $\bar{K}^{st}$ in absolute terms. Given the small magnitude of relativistic interactions observed at the level of the static electronic structure, it is natural to expect that their modulation by nuclear motion ($K$) will also be small. Looking at $\bar{J}^{st}$ more closely, we observe that it is non-zero only for carbon atom displacements in the molecular $xy$ plane. Looking into $\bar{K}^{st}$, we observe the reverse behavior, with derivatives along $z$ ($C_3$ symmetry axis) having the larger contribution.

Figs. \ref{fig3}d-f provide an orthogonal view of the effect of $J$ and $K$ and emphasize which states are more strongly coupled by these contributions to vibronic coupling by averaging over all Cartesian degrees of freedom for specific pairs of states, $ab$. Fig. \ref{fig3}d shows that $\bar{J}_{ab}$ affects states with the same spin multiplicity and $M_S$ to a very large extent, but also couples states with the same $S$ and $\Delta M_S =\pm 2$. This is in agreement with the physical nature of $J$, which has strong spin-preserving contributions plus a small spin-mixing contribution largely introduced by spin-spin coupling (see Fig. S4). Once again, we observe an opposite behavior by looking at $\bar{K}_{ab}$ in Fig. \ref{fig3}e. In this case, the ground-state triplet is coupled to both singlet and triplet excited states and, in all cases, shows a high degree of spin-selectivity for $\Delta M_S=\pm 1$. Fig. S4 shows that $\bar{K}_{ab}$ is to a very large extent determined by spin-orbit coupling. Finally, in Fig. \ref{fig3}, we also report the vibronic coupling among the singlet states and any of the triplets. Notably, all of these transitions are also highly spin selective for $\Delta M_S= \pm 1$, except for those among the first two singlets and the excited triplets. Although all of these vibronic coupling coefficients are estimated at the ground-state equilibrium geometries, we note that spin selectivity is preserved after the optimization of molecular structure on the singlet ground state or the first excited triplet (see Figs. S5 and S6).

Even though vibronic coupling exhibits the same qualitative behavior in \textbf{1} and \textbf{2}, it differs quantitatively in interaction strength. In particular, $J$ is substantially reduced going from \textbf{1} to \textbf{2}. This behavior is confirmed by similar trends for the spin-free vibronic coupling of Eq. \ref{HFT}, as reported in Fig. S7. We advance an interpretation of this finding by recalling that excited-state wavefunctions have a larger multireference character in \textbf{1} than in \textbf{2}, as noted in the previous section. Qualitatively speaking, a wavefunction with a large multireference character is expected to be more likely affected by nuclear displacements. A similar analysis of vibronic coupling is also performed for \textbf{3}, and reported in Figs. S8-S10. For \textbf{3a} we observe a substantial loss of selectivity, likely in response to a reduction of symmetry for the molecular geometry optimized in the crystal. This is partially recovered for the gas-phase optimized structures \textbf{3b-c}.

\begin{figure}[t]
    \centering
    \includegraphics[scale=1]{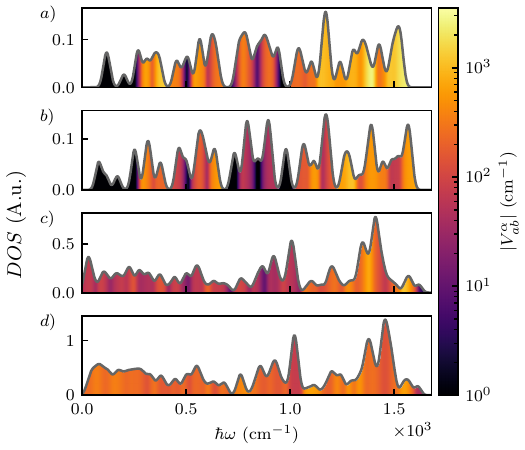}
    \caption{\textbf{Phonons' density of states and vibronic coupling.} Panels a-d) report the phonons' density of states for \textbf{1}, \textbf{2}, \textbf{3c} and \textbf{3a}, respectively. In each case, the color reports the vibronic coupling between any of the three states $T_{0,M_S}$ and any other excited state resolved by phonon frequency.}
    \label{fig4}
\end{figure}

Finally, in Fig. \ref{fig4} we report the vibronic coupling intensity among the ground-state triplet and any excited state resolved by phonons' frequency and phonon density of states. For \textbf{1-2} (Figs. \ref{fig4}a,b) we observe similar features. A visual inspection of phonon modes across different energy windows for \textbf{1} shows that modes at energy lower than 300 cm$^{-1}$ largely correspond to distortions of the molecular plane (see Table S8 and Fig. S11), while modes around 300-500 cm$^{-1}$ include in-plane aromatic breathing modes, as well as out-of-plane hydrogen wagging motions (see Fig. S12). Out-of-plane motions then dominate the vibrational spectrum up until $\sim$ 1000 cm$^{-1}$, where in-plane bending motions involving both hydrogen and carbon atoms take over (see Fig. S13). Carbon-hydrogen stretching motion is finally observed above 3000 cm$^{-1}$. The vibronic coupling strength of these modes can be explained by Figs. \ref{fig3}a-c, which show that distortions along $z$ are much less coupled than in-plane ones. Looking at the same distribution for \textbf{3a} (Fig. \ref{fig4}d), we observe an overall increase in the phonons' density of states at low energy (see also Table S9) and a partial loss of the finer features of vibronic coupling intensity, which now appears more homogeneously spread over different values of energy. To separate the effects of crystal lattice interactions and molecular functionalization, we also report the results for \textbf{3c} in Fig. \ref{fig4}c. The density of states of \textbf{3a} and \textbf{3c} shows similar features above $\sim 350$ cm$^{-1}$, but is markedly different at lower values, pointing to the role of the crystal lattice in shaping the low-energy phonons. Interestingly, however, the lattice structure appears to constrain the very lowest vibrations of \textbf{3c}, which are shifted to higher energy in \textbf{3a}. A visual inspection of the low-energy phonons of \textbf{3a,c} reveals that the corresponding molecular motion is dominated by modes of the bulky tertiary butyl and mesityl groups (see Fig. S14), while at higher temperatures the substituent internal motions are admixed to the triangulene core vibrations, which, as for the gas-phase molecule, are characterized by in-plane motions above $\sim$ 1100 cm$^{-1}$ (see Fig. S15).

\begin{figure}[h!]
    \centering
    \includegraphics[scale=1]{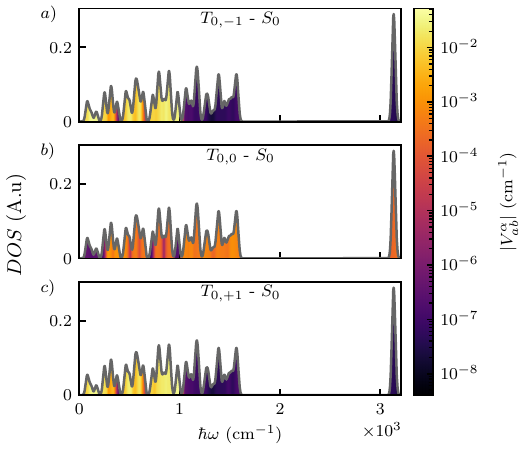}
    \caption{\textbf{Phonons' density of states and spin-resolved vibronic coupling.} Panels a-c) report the phonons' density of states for \textbf{2} color-coded by the intensity of the vibronic coupling between each state $T_{0,M_S}$ and the excited state $S_0$.}
    \label{fig5}
\end{figure}
To conclude the discussion of vibronic coupling, Fig. \ref{fig5} presents the spin-resolved distribution of vibronic coupling for the transitions $T_{0,M_S} \rightarrow S_0 $ in \textbf{2}. Interestingly, the spin-selectivity already emerged in Fig. \ref{fig3} appears intertwined with energy selectivity. As we will show in the following section, the spin selectivity of vibronic coupling has important consequences for inter-system crossing rates. More in detail, we observe a virtually perfect spin selectivity for phonons with energy above $\sim 1000$ cm$^{-1}$, in favor of pairs of states with $M_S=0$, while phonons below about 300 cm$^{-1}$ show the inverse behavior, favoring a coupling among states with $\Delta M_S=\pm 1$. The intermediate range 300-1000 cm$^{-1}$ also shows a high degree of spin selectivity, strongly coupling states with $\Delta M_S=\pm 1$. We can interpret this behavior in terms of molecular angular momentum. The coupling among a triplet and a singlet is due to spin-orbit coupling derivatives, as captured by $K$. In a first-order perturbative effective picture of spin-orbit coupling, we can think of vibronic coupling as $K\propto \sum_i (\partial L_s / \partial X_t) S_s$, where $s$ and $t$ run on Cartesian coordinates. Terms proportional to $S_{x/y}$ are responsible for $\Delta M_S=\pm 1$ transitions, while $S_z$ for $\Delta M_S=0$ ones. With this in mind, for \textbf{1-2} Figs. S16 and S17 clearly show that $L_x$ and $L_y$ are exclusively modulated by displacements along $z$, while $L_z$ is modulated by in-place molecular distortions. This is in perfect agreement with the nature of phonons already discussed. Below 300 cm$^{-1}$ phonons are exclusively displacing atoms along $z$, contributing to $\Delta M_S=\pm 1$ transitions. Above 1000 cm$^{-1}$ modes are exclusively in-plane, selectively coupling $\Delta M_S=0$ states. In the intermediate region, both in-plane and out-of-plane modes exist, making both transitions possible, but still favored for $\Delta M_S=\pm 1$ due to $K_z > K_{x/y}$. Figs. S18-20 reports a similar analysis for \textbf{3a-c}, mirroring previous findings on the partial loss of selectivity.

\begin{figure*}
    \centering
    \includegraphics[scale=1]{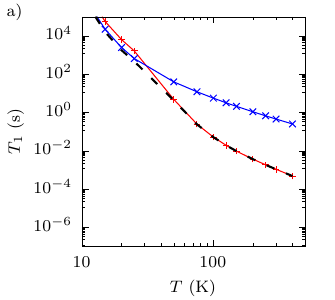}\includegraphics[scale=1]{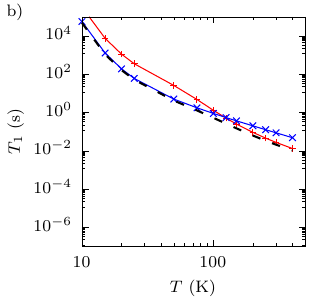}\includegraphics[scale=1]{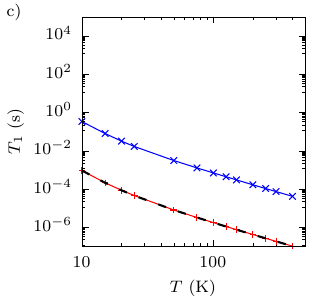} \\
    \includegraphics[scale=1]{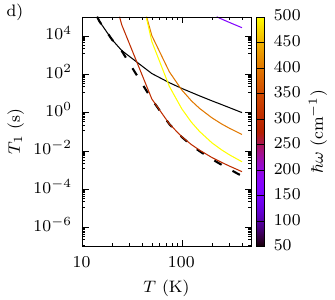}\includegraphics[scale=1]{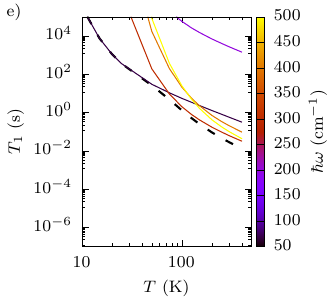}\includegraphics[scale=1]{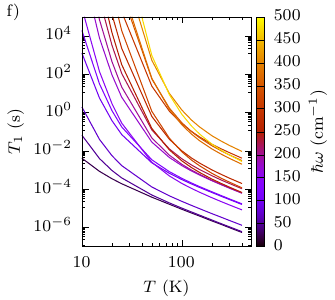}\\
    \includegraphics[scale=1]{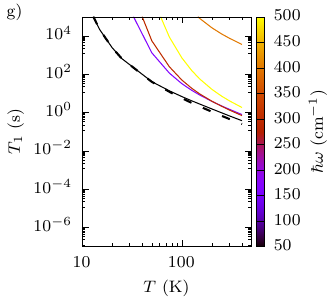}\includegraphics[scale=1]{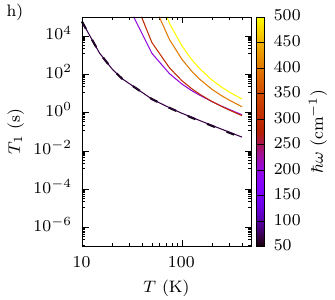}\includegraphics[scale=1]{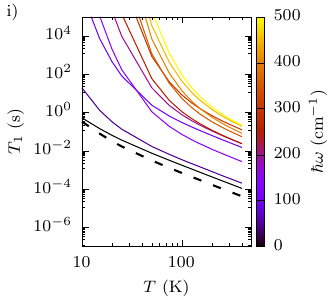}
    \caption{\textbf{Spin-phonon relaxation time as function of temperature.} a-c) The total two-phonon relaxation time, reported as a dashed black line, is decomposed into its linear vibronic coupling (blue line and symbols) and quadratic spin-phonon coupling (red line and symbols) contributions for \textbf{1}, \textbf{2} and \textbf{3a}, respectively. d-f) The total two-phonon relaxation time due to linear vibronic coupling, reported as a dashed black line, is decomposed as a function of phonon envelopes belonging to different energy windows and color-coded according to the sidebar for \textbf{1}, \textbf{2} and \textbf{3a}, respectively. g-i) The total two-phonon relaxation time due to quadratic spin-phonon coupling, reported as a dashed black line, is decomposed as a function of phonon envelopes belonging to different energy windows and color-coded according to the sidebar for \textbf{1}, \textbf{2} and \textbf{3a}. The energy windows width in d-i) are chosen as 100 cm$^{-1}$ for \textbf{1} and \textbf{2} and 25 cm$^{-1}$ for \textbf{3a}, respectively. }
    \label{T1}
\end{figure*}

\textbf{Spin-phonon decoherence.} We now turn to the study of the ground-state triplet spin relaxation. Due to the energy of the molecular electronic excitations, one-phonon resonant processes described by Eq. \ref{W1ph} are neglected. Resonant transitions to excited multiplets are impossible due to the large energy gap between electronic states, and, moreover, the zero-field splitting of the triplet ground state is fractions of a wavenumber, meaning that only the lowest-energy acoustic phonons will be able to resonate with such energy gaps. These transitions would only become relevant at cryogenic temperatures and can be safely neglected in the regime of interest here. We therefore compute the relaxation time, $T_1$, for the $M_S=-1$ component of the ground state triplet as limited by any two-phonon process involving such states, \textit{i.e.} $1/T_1=\sum_{i\ne j}W^{\rm 2-ph}_{ij}$ with $j$ corresponding to the state $T_{0,-1}$. All simulations are carried out at $B=0.33$ T pointing along the $z$ axis. See Fig. S21-S23 for convergence tests. Results for these simulations are reported in Fig. \ref{T1} for \textbf{1-2} and \textbf{3a}. Spin-phonon relaxation times are predicted to be exceptionally long for \textbf{1-2}, with values of $T_1$ of approximately 1 and 27 ms at 300 K, respectively, and their difference well accounted for by the reduction of vibronic coupling in \textbf{2}. This is drastically changed once the molecule is functionalized and included into the crystal, with a $T_1 \sim 0.1$ $\mu$s computed at 300 K for \textbf{3a}. 

Let us now unpack the origin of the values of $T_1$ and its dependence on $T$, starting from splitting $T_1$ into the two possible contributions to two-phonon transitions, namely linear vibronic coupling acting at the fourth-order of perturbation theory (Eq. \ref{eq:5}) and quadratic coupling acting at the second-order of perturbation theory (Eq. \ref{W2phq}). Figs. \ref{T1}a,b show that for \textbf{1-2} the two contributions are similar to one another at low temperature, with quadratic coupling contributions being slightly faster than linear ones. On the other hand, linear vibronic coupling contributions take over closer to 300 K. Interestingly, while quadratic coupling is fully determined by spin-spin interactions, Figs. S25 and S26 show that both spin-orbit and spin-spin interactions can contribute to linear vibronic coupling-limited $T_1$. Turning to the results for \textbf{3a} in Fig. \ref{T1}c, we observe a large dominance of linear vibronic coupling effects over the entire temperature range, although we note that this might be in part due to the approximation taken in computing quadratic coupling for the functionalized molecules. Quadratic coupling has indeed been neglected for the atoms belonging to the peripheral groups of \textbf{3} to reduce the number of DFT calculations needed from 290,322 to 20,808. 

We now move on to identifying the phonons responsible for the various relaxation mechanisms. Figs. \ref{T1}g-i show that quadratic-coupling contributions to $T_1$ are fully determined by the lowest-energy phonons at any temperature. This is in agreement with previous observations for transition metal complexes\cite{Garlatti2023TheQubitsb} and is explained by the absorption-emission two-phonon process depending on the Bose-Einstein population of one of the two modes, thus favoring the lowest-energy phonons. On the other hand, linear vibronic coupling contributions to $T_1$ exhibit richer behavior relative to the energy of the phonons involved. Figs. \ref{T1}d-f show that two main regimes are observed for \textbf{1-2}, with low energy phonons determining relaxation at low-$T$, while phonons with energies in the range of 300-500 cm$^{-1}$ become activated toward room-$T$. This is particularly dramatic in \textbf{1}. Despite these transitions depending on the Bose-Einstein population exactly as for quadratic coupling, virtual excitations to high-energy electronic states partially balance out population effects. For \textbf{3a} we can now appreciate the dominant effect of low-energy phonons, which, as shown in Fig. \ref{fig4}c, have a much larger density of states than \textbf{1-2}.

The contribution of linear vibronic coupling to $T_1$ has also been computed for \textbf{3b,c}, and it is reported in Fig. S26, where it is possible to appreciate that a smaller envelope of low-energy phonons than in \textbf{3} is responsible for limiting $T_1$. A visual inspection in Fig. S14 confirms the key role of the functionalization groups.

To fully unravel the mechanism underpinning $T_1$ for linear vibronic coupling, we also identify the virtual transition paths that lead to relaxation. While multiple paths can contribute depending on the molecule and the temperature range considered, we can identify some underlying principles based on the analysis of vibronic coupling presented in Fig. \ref{fig3}. Virtual transition paths will likely follow the steps with the largest vibronic coupling. For instance, starting from a triplet state with $M_S=-1$, the strongest virtual transition is to excited triplet states with the same $M_S$. Once virtually arrived at this state, the next possible relaxation pathway goes down to $M_S=1$, which is enabled by spin-mixing effects in $J$ introduced by spin-spin interactions. Another possible virtual path would instead involve reaching the ground-state triplet with $M_S=0$ from the excited triplet, this time largely enabled by spin-orbit coupling contributions to $K$. Singlet excited states could be expected to play a smaller role in virtue of the fact that all virtual transitions involving them would necessarily include only $K$ terms, but still contribute at low $T$ due to their lower energy with respect to the triplets. 

Finally, we compute the vibrational pure dephasing contribution to $T_2$. Results show that whilst a pure dephasing contribution is visible, $T_2$ remains of the same order of magnitude as $T_1$, e.g. at 300 K, $T_2=1.9$ and $T_2=13.3$ ms in \textbf{1} and \textbf{2}, and $T_2=0.2$ $\mu$s in \textbf{3}.

\textbf{Spin-spin dephasing.} The CCE method is used to simulate the decoherence profile of the triplet ground state under the effect of nuclear spin dynamics. For these simulations, a magnetic field of $B=0.33$ T is applied along the molecular symmetry axis perpendicular to the molecular plane, to resemble a conventional X-band EPR experiment. The full set of convergence tests for these simulations are provided as ESI.

\begin{figure*}[t]
    \centering
    \includegraphics[scale=1]{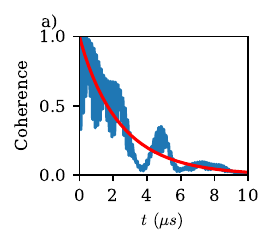} \includegraphics[scale=1]{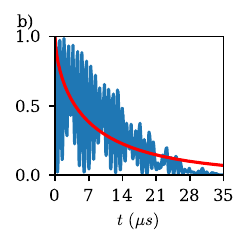}
    \includegraphics[scale=1]{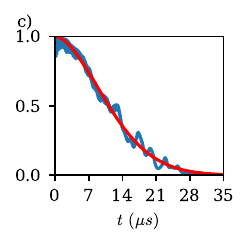}
    \includegraphics[scale=1]{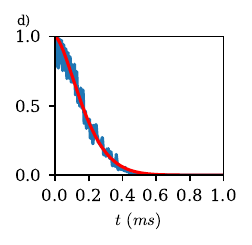}
    \caption{\textbf{Hahn-echo profile in 3.} Decoherence profile of the electron spin of \textbf{3} for a Hahn-echo experiment in a bath of a) hydrogen nuclear spins with a coherence time of $T_2$=2.42 $\mu$s and a stretch factor of $\beta$=0.95, b) deuterium nuclear spins with hydrogen spins on the central molecule with a coherence time of $T_2$=7.49 $\mu$s and a stretch factor of $\beta$=0.63, c) hydrogen spins only on the central molecule with a coherence time of $T_2$=13.69 $\mu$s and a stretch factor of $\beta$=1.79, and d) deuterium spins only on the central molecule with a coherence time of $T_2$=0.21 ms and a stretch factor of $\beta$=1.79.}
    \label{T2}
\end{figure*}

Due to the delocalized nature of the spin in these molecules, the hyperfine interaction with hydrogen spins in the central molecule includes the isotropic Fermi contact, dipole-dipole, and orbital contributions. Simulations of Hahn-echo decay are performed considering all possible initial coherent superpositions among two of the three states of the ground-state triplet multiplet. The full set of details for all subsequent simulations is provided in the ESI file.  Fitting each of the decoherence profiles seen in Fig.~\ref{T2} with a stretched exponential decay,
\begin{equation}\label{st_exp}
    \mathcal{L}(t) = e^{-(t/T_2)^\beta},
\end{equation}
the complete set of simulated coherence times and stretch factors is collated in Table~\ref{Decoherence table}.

\begin{table}[h!]
    \begin{center}
    \caption{\textbf{Spin-spin decoherence in 1-3.} Simulated $T_2$ and stretch factors $\beta$ for \textbf{1-3} under a Hahn-echo sequence in different nuclear magnetic environments. X-H (X-D) points to the use of Hydrogen (Deuterium) as a nuclear isotope.}
    \label{Decoherence table}
     \begin{tabular}{ c|c|c|c }
     \hline
     \hline
     Central molecule & Molecular environment & $T_2$ ($\mu$s) & $\beta$ \\
     \hline
     \textbf{1}-H & - & 44.38 & 1.98 \\
     \textbf{1}-D & - & $\infty$ & - \\
     \textbf{2}-H & - & 43.76 & 1.96 \\
     \textbf{2}-D & - & $\infty$ & - \\
     \textbf{3a}-H & \textbf{3a}-H & 2.42 & 0.95 \\
     \textbf{3a}-H & \textbf{3a}-D & 7.49 & 0.63 \\
     \textbf{3a}-H & - & 13.69 & 1.79\\
     \textbf{3a}-D & - & 210 & 1.79\\
     \hline
     \hline
    \end{tabular}
    \end{center}
\end{table}

We start by noting that simulations for \textbf{1-2} return the longest coherence times in the case of the hydrogen spin bath environment, while there is a complete absence of decoherence in the deuterated molecule. This is to be expected, as both \textbf{1} and \textbf{2} are single-molecule systems. In the hydrogen spin environment, the 12 hydrogen spins present in the environment are sufficiently many and strongly interacting for the central spin to fully decohere (see Figs. S36 and S39). For the deuterated environments, results show that the central spins of \textbf{1} and \textbf{2} show no level of decoherence over timescales of 1 s  (see Figs. S37 and S40). In this case, the deuterium spin environment has neither sufficiently many spins, nor sufficiently strong interactions with the central spins to cause any meaningful decoherence. 

We then move on to the simulation of spin dephasing rates due to nuclear spin dynamics in a more realistic environment, as allowed by the knowledge of the crystalline structure of \textbf{3}. The first case under study is the dephasing of the central electronic spin as a result of spin-spin dipolar interactions with the hydrogen spins of both the central molecule and nearby molecules in the crystal. The interaction with hydrogen spins external to the central molecule is computed from the classical interaction among point dipoles. The decoherence profile for this case can be seen in Fig.~\ref{T2}a, where fitting with Eq.~\ref{st_exp} produces a coherence time of $T_2=2.42$ $\mu$s when initializing the spin in a coherent superposition of the $T_{0,-1}$ and $T_{0,0}$ spin levels. This value is consistent with expectations, as \textbf{3} has a more concentrated hydrogen environment ($\sim 0.06$ hydrogen spins per {\AA}$^3$) compared to other TM-based spin-1/2 molecular crystals recently studied with the same method ($\sim 0.03$ hydrogen spins per {\AA}$^3$) and exhibiting values of $T_2$ of about 10 $\mu$s.\cite{Ryan2025SpinBaths} Results for the other superpositions of the ground state triplet spin levels are provided as ESI. 

Next, we investigate the origin of such a value for $T_2$ and possible strategies to enhance it. A commonly employed experimental technique to reduce dephasing in organic lattices is to exchange hydrogen spins outside of the central molecule with deuterium spins, which carry a weaker gyromagnetic ratio ($\gamma_H=26.7522$ rad$\cdot$kHz$\cdot$G$^{-1}$ vs $\gamma_D=4.1065$ rad$\cdot$kHz$\cdot$G$^{-1}$) and therefore participate more weakly in spin-spin dipole interactions. This is generally accomplished by dissolving the molecule in a deuterated solvent, but we here employ the native crystal structure of \textbf{3} to achieve a direct comparison with the fully hydrogenated case. Fig.~\ref{T2}b shows the decoherence profile produced under these conditions, which now provides a $T_2$ value of 7.49 $\mu$s upon fitting with Eq.~\ref{st_exp}.

Interestingly, in this case, the deuterium spins from other molecules around the crystal contribute to the bottleneck for $T_2$. This can be seen from Fig.~\ref{T2}c, in which all spins are neglected except for the hydrogen spins on the central molecule. In this case, we obtain $T_2=13.69$ $\mu$s, exceeding the coherence time once the deuterium spins are added. Behavior such as this suggests an influential role for the spin diffusion barrier in this decoherence process. A discussion of the results for the fully deuterated system is provided as ESI. Pushing this analysis even further, if the central molecule is itself deuterated, and the local magnetic environment of the central spin is designed in such a manner as to be spin-free\cite{Zadrozny2015MillisecondQubit}, $T_2$ increases further by over one order of magnitude to 0.21 ms, as seen in Fig.~\ref{T2}d, offering a chemical way to achieve a very high ceiling for this figure of merit.  

\textbf{Excited states relaxation.} We now explore excited-state relaxation processes. Although we postpone the full determination of electronic excited states dynamics to a future study, we here provide a semi-quantitative prediction of some possible relaxation pathways and discuss their spin selectivity and molecular origin. For this task, we focus on those key events potentially leading to ODMR and spin initialization: i) radiative rates from the optically accessible $T_1$ state, and ii) the two supposedly leading ISC channels, i.e. $T_1$ to $S_2$ and $S_0$ to $T_0$. To this end, we relax the molecular geometry of \textbf{1} and \textbf{2} for the states $T_1$ and $S_0$. We note that the DFT-based optimization of excited states geometries in the presence of degeneracies and multireference character is not free of challenges. In particular, we observe a general disagreement between DFT and multireference methods on the nature of the singlet ground state at distorted geometries (see discussion in ESI). We argue that multireference methods offer a higher reliability and consider the closed-shell singlet for our $S_0$ geometry optimization.

Starting from radiative processes, we estimate the radiative lifetime of the excited triplets, $\tau_r$, through the expression given by the Wigner-Weisskopf theory of fluorescence,\cite{Gali2019AbDiamond, Weisskopf1930BerechnungLichttheorie} which reads as
\begin{equation}
    \frac{1}{\tau_r} = \frac{n_r \omega^3 |e\boldsymbol{r}|^2}{3 \pi \varepsilon_0 \hbar c^3} \:,
\end{equation}
where the refractive index, $n_r$, is set to 1 for a molecule, $\hbar \omega$ is the transition energy, and $e\boldsymbol{r}$ is the electric transition dipole moment. The radiative rates computed at the optimized first excited triplet geometry are $1.94 \: 10^6$ s$^{-1}$ and $1.86 \: 10^6$ s$^{-1}$ for \textbf{1} and \textbf{2}, respectively, and in good agreement with the experimental value of $2.7  \: 10^6$ s$^{-1}$ for a functionalized core of \textbf{2} in various solvents\cite{Arikawa2023AState}. This result is particularly interesting in light of the large difference in oscillator strengths for the two molecules, which are computed at the spin-free level as $f=5.2 \:  10^{-3}$ and $f=1.5 \: 10^{-2}$, respectively. This is explained by the reduced triplet-triplet gap in \textbf{2}, which partially offsets the benefit of a large oscillator strength. Assuming that the overall radiative rate is not dramatically affected by side-band contributions introduced by Franck-Condon effects, we interpret the experimental evidence of luminescence exclusively for \textbf{2} due to a partial quenching of its non-radiative relaxation mechanisms. This is also supported by the computed reduction of vibronic coupling in \textbf{2}.

We then discuss some of the possible non-radiative relaxation pathways for \textbf{2}, on account of being the core of a known luminescent triangulene derivative\cite{Arikawa2023AState}. Within the formalism employed here, Franck-Condon and Herzberg-Teller factors arising from molecular geometry reorganization are neglected.\cite{Hagai2026ExtendedMaterials} This effectively means that no vibronic first-order transition involving multiple quanta of the same mode can be invoked to match energy gaps exceeding the phonons' spectral range. However, we note that i) \textbf{2} presents gaps of $\sim$3677 cm$^{-1}$ between $T_0$ and $S_{0/1}$ and $\sim$2845 cm$^{-1}$ between $S_2$ and $T_{1/2}$,  allowing two-phonon resonant contributions to ISC rates within our current formalism, and ii) spin-orbit coupling does not directly mix start/end states of ISC, making direct processes generally less favored than two-phonon ones. We first report ISC rates computed with Eq. \ref{eq:5} at the ground-state geometry for pedagogical reasons, and then account for the effect of the excited states reorganization for a proper estimation of ISC rates. Table \ref{ISCtab} reports the transition rate among singlet and triplet states computed within the exact same formalism of Eq. \ref{W2phq} used for simulating spin-phonon relaxation.

\begin{table}[h!]
    \begin{center}
    \caption{\textbf{Spin-resolved ISC rates at 300 K for \textbf{2}.} The two-phonon rates $W_{fi}$ from the initial state $i$ to the final state $f$ are predominantly mediated by virtual transitions to an intermediate state $c$ due to the absorption/emission of two phonons with energy $\hbar\omega_\alpha$ and $\hbar\omega_\beta$. Negative (positive) energies correspond to a phonon emission (absorption). Relevant modes for ISC at $S_0$ and $T_1$ geometries are imaged in Figs. S42 and S43.}
    \label{ISCtab}
     \begin{tabular}{ c|c|c|c|c|c|c }      
    \multicolumn{7}{c}{} \\
    \multicolumn{7}{c}{Ground-state optimized geometry} \\
     \hline
     \hline
     State $i$ & $S_{0}$ & $S_{0}$ & $S_{0}$ & $T_{1,-1}$ & $T_{1,0}$ & $T_{1,-1}$ \\
     State $f$ & $T_{0,-1}$ & $T_{0,0}$ & $T_{0,1}$ & $S_{2}$ & $S_{2}$ & $S_{2}$  \\
     $W_{fi}$ (s$^{-1}$) & $5.4 \: 10^{3}$ & $1.3 \: 10^{3}$ & $7.0 \: 10^{3}$ & $3.3 \: 10^{1}$& $1.0 \: 10^{5}$ & $4.5 \: 10^{1}$ \\
     State $c$  & $T_{1/2,-1}$ & $T_{1/2,0}$ & $T_{1/2,1}$ & $S_{0/1}$ & $S_{0/1}$ & $S_{0/1}$ \\
     $\hbar\omega_\alpha$ (cm$^{-1}$) & -550 & -566 & -550 & 250 & -1383 & 309 \\
     $\hbar\omega_\beta$ (cm$^{-1}$) & -3130 & -3130 & -3127 & -3129 & -1445 & -3145 \\
     \hline
     \hline
    \multicolumn{7}{c}{} \\
     \multicolumn{7}{c}{Excited-state optimized geometries} \\
     \hline
     \hline
     State $i$ & $S_{0}$ & $S_{0}$ & $S_{0}$ & $T_{1,-1}$ & $T_{1,0}$ & $T_{1,1}$ \\
     State $f$ & $T_{0,-1}$ & $T_{0,0}$ & $T_{0,1}$ & $S_{2}$ & $S_{2}$ & $S_{2}$  \\
     $W_{fi}$ (s$^{-1}$) & $8.0 \: 10^{3}$ & $6.4 \: 10^{-1}$ & $8.2 \: 10^{3}$ & $1.7 \: 10^{6}$& $7.4 \: 10^{4}$ & $3.7 \: 10^{6}$ \\
     State $c$  & $T_{1/2,-1}$ & $S_{1}$ & $T_{1/2,1}$ & $S_{0/1}$ & $S_{0/1}$ & $S_{0/1}$ \\
      $\hbar\omega_\alpha$ (cm$^{-1}$) & -229 & -291 & -229 & 72 & -316 & -86 \\
     $\hbar\omega_\beta$ (cm$^{-1}$) & -3127 & -3127 & -3127 & -1550 & -1183 & -1420 \\
     \hline
     \hline
    \end{tabular}
    \end{center}
\end{table}

\begin{figure}[h!]
    \centering
    \includegraphics[scale=1]{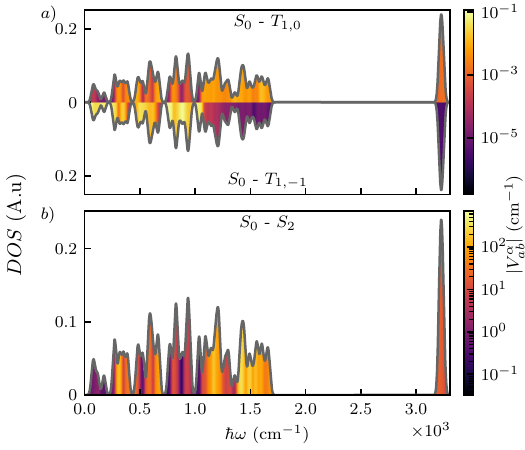}
    \caption{\textbf{Vibronic coupling involved in ISC processes for 2.} The phonons' density of states computed at the $T_1$ optimized geometry of \textbf{2} is color-coded by the vibronic coupling intensity for the relevant pairs of states involved in the virtual processes leading to ISC: a) $S_0 - T_1$, and b) $S_0 - S_2$.}
    \label{fig8}
\end{figure}

Starting from the case of ISC rates computed at the ground-state geometry, we observe two interesting facts: i) ISC from $T_1$ to $S_2$ is highly spin conserving, with the transition from the $M_S=0$ component of the triplet predicted four orders of magnitude more efficient than for $M_S=\pm 1$, while ii) the reverse ISC from $S_0$ to $T_0$ exhibits the opposite spin selectivity, by slightly favoring the conversion of $M_S=0$ into $M_S=\pm 1$. To understand this result, we decompose the contributions to $T_{1/2} \rightarrow S_2$ ISC and find that virtual transitions involving $S_0$ and $S_1$ intermediate states are mediated by two-phonon processes merging the total energy gap $\sim$ 2800 cm$^{-1}$. For the $M_S=0$ component, the double emission of a phonon with energy of $\sim 1400$ cm$^{-1}$ is the preferred mechanism, due to involving phonons strongly coupled to both steps of the virtual transition, i.e. $T_{1/2} \rightarrow S_{0/1}$ and $S_{0/1} \rightarrow S_2$. This can be appreciated in Figs. \ref{fig8}a,b. On the other hand, for the $M_S=\pm 1$ components, this pathway is not feasible, due to the vanishing vibronic coupling for the first step of the virtual transition $T_{1/2} \rightarrow S_{0/1}$ (see Fig. \ref{fig8}a). For this transition, the preferred pathway involves the absorption of a phonon of 250-300 cm$^{-1}$ for the virtual process $T_{1/2} \rightarrow S_{0/1}$ and the emission of a phonon with about 3100 cm$^{-1}$ for $S_{0/1} \rightarrow S_2$, both showing finite coupling. Interestingly, this synergy of ss-ISC events is exactly what would be needed to generate both ODMR contrast and spin initialization in favor of $M_S=\pm 1$ states. Indeed, the larger ISC rate computed for the $T_{1/2,0}$ state would dimish its quantum yield, making $M_S=\pm 1$ states brighter. At the same time, the $M_S=0$ population undergoing ISC would be funneled into $M_S=\pm 1$ states by the second ISC event, increasing their population towards spin initialization.

We then explore the role of excited states' energy reorganization by recomputing the rates of the two ISC events after the molecular geometry of \textbf{2} has been optimized for the first excited triplet and the ground-state singlet, respectively. For the $T_{1} \rightarrow S_{2}$ ISC process, we observe a drastic change of spin selectivity, now favoring the $M_S=\pm 1$ components by two orders of magnitude. This is explained by the lowering of the energy gap between the states involved, now separated by only 1400 cm$^{-1}$. Indeed, although virtual transitions to $S_{0/1}$ states are still mediating the process, the two phonons involved for the $M_S=\pm 1$ states are now different and involve one low energy mode at $\sim 70-90$ cm$^{-1}$ and the emission of a second one at $1400-1500$ cm$^{-1}$. According to Figs. \ref{fig8}a,b, this is the most favorable combination of two phonons for these virtual processes, indeed resulting in quite fast ISC rates. The exact same pathway is not favorable for the $M_S=0$ component due to the vanishing coupling of the lowest energy phonons to the transition $T_{1/2} \rightarrow S_{0/1}$, which instead requires the emission of a phonon of energy $\sim 300$ cm$^{-1}$, leaving the emission of a second phonon of $\sim 1200$ cm$^{-1}$ to mediate the virtual process $S_{0/1} \rightarrow S_2$. Both phonons have a finite, but non-optimal, coupling, leading to a slower ISC rate. This flavor of ISC closely mimics the one of NV$^{-}$ centers, and would support the presence of ODMR in favor of $M_S=0$ states. This process is now also amenable to single-phonon direct resonant transitions due to the reduced $T_1 - S_2$ energy gap. We find that direct processes have a vanishingly low rate for $\Delta M_S=\pm 1$ transitions and a commensurate one to two-phonon rates for $\Delta M_S=0$ ($ W^{\rm 1-ph} = 5.7 \: 10^{4}$), supporting the robustness of our conclusion based on two-phonon processes.

Relative to the ground-state geometry, the use of $S_0$ structure for the calculation of the second ISC event leads to a much stronger spin selectivity in favor of the $M_S=\pm 1$ states. The transitions with $\Delta M_S=\pm 1 $ can be accommodated both in terms of vibronic coupling intensity and energy conservation by emitting a phonon of energy $\sim 230$ cm$^{-1}$ during the virtual transition $S_0 \rightarrow T_{1/2,\pm 1}$ and emitting a second phonon of $\sim 3127$ cm$^{-1}$ for the virtual transition $T_{1/2,\pm 1} \rightarrow T_{0,\pm 1}$. The first virtual transition cannot be accommodated in the case of $\Delta M_S=0$ due to their vanishing coupling with phonons of energy lower than 300 cm$^{-1}$, and a virtual process involving the $S_1$ state, now not degenerate anymore with $S_0$, and a phonon at 291 cm$^{-1}$ becomes the preferred, yet very inefficient, relaxation pathway. 

This scenario potentially lifts the benefit of a strongly ss-ISC between $T_{1}$ and $S_2$ and might impede the accumulation of spin in a given subset of levels of the ground-state manifold. However, an important observation emerges from our analysis: the flavor of spin-selectivity of ISC can be highly modulated by engineering the states' energies. Indeed, for even a slightly larger renormalization energy of 400 cm$^{-1}$ for $S_{0/1}$, the spin selectivity would be reversed in favor of $\Delta M=0$, which could then lead to a strong spin initialization into $M_S=0$, akin to the optical processes in NV$^{-}$ centers.

To conclude, we note that, even in the event of multi-quantum direct processes overcoming two-phonon resonant ones considered here, the spin selectivity of ISC would likely remain very high, with the possibility of selecting its flavor by engineering the energy of the states involved. Indeed, as Fig. S44 shows, vibronic coupling among $T_{1/2} \rightarrow S_2$ $S_0 \rightarrow T_0$ differs between $\Delta M_S=0$ and $\Delta M_S=\pm 1$ at all frequencies. 

\section*{Discussion and Conclusions}

The development of molecular color centers able to replicate the spin and photophysical properties of NV$^{-}$ centers requires a careful fine tuning of their properties to achieve three key conditions: 1) A highly coherent triplet ground state well separated from any excited state ($\Delta E \gg \mathrm{k_B}T$), 2) bright, fast relaxing, excited states, and 3) highly spin selective intersystem crossing dimming a single spin channel and funneling population from excited states into a single $M_S$ component of the ground state. While condition 1 allows to preserve and manipulate quantum information, conditions 2 enables efficient optical addressing on timescales shorter than decoherence, and condition 3 provides a means to read out the ground state through ODMR as well as initialize its population in a pure state through optical spin pumping. 

In this contribution, we have provided theoretical evidence that idealized triangulene-based diradicals can fulfil such conditions. Both \textbf{1} and \textbf{2} exhibit an electronic structure strikingly similar to NV$^{-}$ centers, including a well-isolated isotropic triplet ground state. As a consequence of this electronic structure lacking low-lying excited states, and in virtue of the high rigidity of the triangulene organic scaffold, the predicted $T_1$ for these molecular templates exceed the ms timescale. If realized, such long $T_1$ would compete with those of deeply buried NV$^{-}$ centers.\cite{Jarmola2012Temperature-Diamond} Regarding condition 2, molecule \textbf{2} is modeled after the core of a known luminescent stable triangulene diradical with quantum yield $\sim 1 \%$ and radiative lifetime of $\sim 10^{-6}$ s, metrics which would allow for optical addressing over timescales shorter than $T_2$, especially in the case of a deuterated \textbf{2}. Regarding condition 3, our detailed study of vibronic coupling and ISC allowed us to unearth the high level of spin selectivity that these processes exhibit in triangulene diradicals. Depending on the nuances of excited-state geometry reorganization, ISC can favor $M_S=\pm 1$ or $M_S=0$ components, but in all studied instances, a high level of spin selectivity is observed for the bright excited triplets, supporting the potential to observe ODMR signal, as well as spin initialization upon fine-tuning the energy levels.  

On the one hand, the study of these idealized diradicals supports their potential for quantum technology. On the other hand, the study of \textbf{3}, i.e. a real crystal of \textbf{1}, highlights the challenges in delivering the performance of ideal systems in practice. Both the functionalization and the introduction of a lattice introduce low-energy phonons detrimental to $T_1$, which drops by several orders of magnitude. In these conditions, $T_2$ becomes limited to 0.21 ms at 10 K for a hypothetical deuterated molecule of \textbf{3} in a nuclear spin-free environment like a frozen solution of CS$_2$.\cite{Zadrozny2015MillisecondQubit} Interestingly, ab initio simulations points to the specific molecular motion of the tBu and mesityl groups as responsible for this deterioration of $T_1$. This suggests possible ways through, for instance by replacing the groups with bulky but rigid groups, or by integrating the core structure of \textbf{1-2} into rigid covalent organic frameworks (COFs), as recently shown possible for some triangulene-based systems.\cite{Lakshmi2020AStates} We note that such a proposed search for the long $T_1$ limit imposed by the isolated molecular structure might be particularly fruitful for these rigid organic compounds. Indeed, while low-energy vibrations have already been identified as limiting $T_1$ in spin-1/2 molecular qubit prototypes based on transition metals (TMs),\cite{Garlatti2023TheQubitsb} the $T_1$ of such compounds would likely remain limited by low-frequency intra-molecular motion also in the absence of a lattice. We confirm this here by computing a $T_1$ for the CrN(pyrdtc)$_2$ in gas phase, which shows only a marginal improvement of $T_1$ with respect to the presence of a solid-state lattice\cite{Mariano2025TheMolecules} (see Fig. S45). Recent literature on integrating TM-based spin-1/2 into metal-organic frameworks\cite{Suzuki2026SpinlatticeMaterials} versus integrating organic radicals into COFs\cite{doi:10.1021/jacs.5c09638} appears to also support such observations. In addition to suppressing the vibrational noise originating from low-energy molecular and lattice dynamics, the intrinsic luminescent properties of our prototype molecule \textbf{2} would also require further optimization for this class of organic diradical to emerge as an ideal molecular quantum platform. However, since the experimental quantum yield of 1\% was measured on a crystal of \textbf{2} with a functionalization equivalent to \textbf{3},\cite{Arikawa2023AState} we anticipate that the vibrational control necessary to increase $T_1$ might also improve quantum yield. Additionally, alternative ways to break alternant conjugation to favor fast emissive rates should also be explored. For instance, while alternant diradicals suffer from small triplet-singlet gaps ($\sim \mu$eV), they have been shown to achieve $\sim 100\%$ quantum yield once properly functionalized.\cite{Chowdhury2025BrightStates} Similar strategies could be explored here in view of producing similar effects.

The chemical functionalization of the scaffold of \textbf{1-2} clearly stands out as the key next step in the realization of our proposal to use them as optically addressable qubits with long coherence times. Luckily for us, these molecules are underpinned by a vast chemical space for chemists to play with.\cite{doi:10.1021/acs.chemrev.3c00406} Indeed, triangulene itself can be tailored to fine tune its electronic and vibrational structure. In this study itself, we have shown how the N$^{+}$ substitution in \textbf{2} has dramatic effects in reducing vibronic coupling by breaking alternant conjugation. Similarly, the comparison between \textbf{1} and \textbf{3} shows how the chemical functionalization of the core has the many-fold role of altering energy level spacings, changing the low-energy vibrational structure and improving chemical stability. These results point to a high degree of tunability that can be exploited to enhance specific properties of these diradicals. Going beyond simple triangulene, these molecules represent the smallest size member of the nanographenes family,\cite{Zeng2021Open-ShellFragments} where size, topology, and multi-center exchange can be further exploited as fine-tuning degrees of freedom to achieve the desired quantum properties. Importantly, our in-depth ab initio analysis points to the spin-selectivity of vibronic coupling as a general property of $\pi$ conjugated systems. Indeed, it stems from the interplay between in/out-of-plane vibrations and how they selectively affect molecular angular momentum. The energy distribution of spin-resolved vibronic coupling is also likely to hold for other planar carbon structures, including arenes and nano-graphenes, due to the general consistency of the energy of specific in-plane/out-of-plane vibrational modes.

Given the theoretical nature of the work, we would also like to comment on the technical aspects of our methodology. Here, we have provided the first comprehensive ab initio characterization of spin decoherence in a realistic solid-state environment, inclusive of both spin relaxation rates computed from relativistic vibronic coupling and spin decoherence through cluster expansion and quantum master equations, for spin-spin and spin-phonon interactions, respectively. On top of this, we have moved a step towards the integration of excited-state dynamics into this ab initio framework. To the best of our knowledge, this work marks the first attempt to quantitatively simulate spin-resolved ISC rates in molecules within a fully ab initio approach. While the journey towards a fully reliable prediction of spin-resolved ISC has just started and the implementation of excited PES orthogonality breaking is already ongoing, the leading ISC rates are found to be consistent with those reported in literature for organic molecules, including when two-phonon processes are accounted for.\cite{Hagai2026ExtendedMaterials} A key feature of ISC that emerges from simulations is the sensitivity of rates on energy levels alignment. While placing a high bar in terms of the accuracy required of ab initio simulations, this observation supports the possibility of chemically tuning ISC efficiency and represents the perfect case for the deployment of advanced computational methods for molecular design. More importantly, our study reflects the importance of establishing a quantitative, fully ab initio theory of spin-resolved electronic relaxation. Indeed, the nature of the intricate multi-phonon vibronic pathways and the interplay among spin and energy selectivity in ISC and relaxation observed here could only have been unravelled through an ab initio approach that goes beyond equilibrium static considerations, and further work in this direction should be a priority for this nascent field.

In conclusion, we have presented the first application of first-principles spin-phonon relaxation theory to organic (di)radicals, as well as taken a step towards unifying the theory of spin and excited-state dynamics within the same first-principles formalism. In applying these methods to triangulene-based diradicals we have shown that this family of compounds represents a very promising and versatile playground for the development of next-generation optically-active molecular quantum technologies, as well as extracted novel design rules on how to control molecular spin coherence and optical addressability in high-spin conjugated systems for future exploration.

\vspace{0.2cm}
\noindent
\textbf{Acknowledgements and Funding}\\
Computational resources were provided by Trinity College Dublin Research IT, the Irish Centre for High-End Computing (ICHEC), and EuroHPC JU. A.L. acknowledges the funding from the European Research Council (ERC) (grant agreement No. 948493). A.S. acknowledges the funding through the Marie Sklodowska Curie Action 101151501. C.R. and C.H. acknowledge funding from Taighde Éireann - Research Ireland through the Government of Ireland Postgraduate Scholarship (project IDs GOIPG/2024/4498 and GOIPG/2024/5473).

\vspace{0.2cm}
\noindent
\textbf{Author Contributions}\\
A.L. conceptualized and coordinated the project. A.S. ran the ab initio simulations with contributions from C.H. and L.A.M. C.H. and L.A.M. contributed to the software for the simulation of vibronic coupling. C.R. ran the spin-spin decoherence simulations. A.L. ran the simulation of spin-phonon decoherence and ISC rates, interpreted the results of simulations, and wrote the manuscript with contributions from all authors.

\vspace{0.2cm}
\noindent
\textbf{Conflict of Interest}\\
The authors have no competing interests to declare.



\end{document}